# Automated HEMT Model Construction from Datasheets via Multi-Modal Intelligence and Prior-Knowledge-Free Optimization

Yuang Peng, Jiarui Zhong, Yang Zhang, Hong Cai Chen, *Member, IEEE*

*Abstract*—Parameter extraction for industry-standard device models like ASM-HEMT is crucial in circuit design workflows. However, many manufacturers do not provide such models, leaving users to build them using only datasheets. Unfortunately, datasheets lack sufficient information for standard step-by-step extraction. Moreover, manual data extraction from datasheets is highly time-consuming, and the absence of a fully automated method forces engineers to perform tedious manual work. To address this challenge, this paper introduces a novel, end-to-end framework that fully automates the generation of simulation-ready ASM-HEMT SPICE models directly from PDF datasheets. Our framework is founded on two core innovations: 1) a multi-modal AI pipeline that integrates computer vision with a large language model (LLM) to robustly parse heterogeneous datasheet layouts and digitize characteristic curves, and 2) a novel Iterative-Focusing Tree-structured Parzen Estimator (IF-TPE) optimization algorithm is specifically designed for device parameter extraction under the high-dimensional, sparse-data condition by adaptively refining the parameter search space. Experimental validation on a diverse set of 17 commercial HEMT devices from 10 manufacturers confirms the framework's accuracy and robustness. The generated models demonstrate excellent agreement with published DC and RF characteristics. As the first fully automated workflow of its kind, our proposed solution offers a transformative approach to device modeling, poised to significantly accelerate the circuit design cycle by eliminating the need for manual parameter extraction.

## I. INTRODUCTION

HIGH-electron-mobility transistors (HEMTs) have become indispensable components in radio frequency (RF) and radar systems owing to their superior power density and high-frequency performance[1], [2]. Accurate device modeling is paramount for circuit design and performance evaluation. Methodologies for HEMT modeling are broadly categorized into three types: physics-based models [3], [4], [5], which are computationally prohibitive; equivalent circuit models [6], [7], [8], which represent the prevailing paradigm; and data-driven models [9], [10], [11], which necessitate extensive datasets for training.

The Advanced SPICE Model for HEMTs (ASM-HEMT) [12] is the industry-standard for modeling Gallium Nitride (GaN) HEMTs, particularly for RF circuit design. A significant challenge, however, is its complex parameter extraction. Traditionally, this process requires extensive data obtained from specialized, high-cost measurement systems [13]. Consequently, limited access to these systems has made public device datasheets the primary information source for many engineers and researchers.

Nevertheless, deriving an accurate ASM-HEMT model solely from a datasheet presents two fundamental challenges [14],[15]: 1) **data incompleteness**, as datasheets typically provide key I-V and S-parameter curves but omit essential geometric and intrinsic physical parameters required by the model; and 2) **laborious data extraction**, manually digitizing information from unstructured PDF documents is labor-intensive, time-consuming, and error-prone. Consequently, a fully automated methodology for generating simulation-ready ASM-HEMT models directly from datasheet information has not yet been reported in the literature.

To address these challenges, the first critical step is to achieve automated datasheet parsing. Previous works have addressed single-modal extraction of either text [16] or graphical plots [17], but lacked an integrated approach. The advent of large language models (LLMs) has enabled parsing of design specifications from text [18] and qualitative descriptions of charts [19], yet these methods fall short of producing the precise quantitative data required for simulation. While multi-modal frameworks like DocEDA [20] and ModelGen [21] have emerged, they primarily function as assistive tools and lack the robustness to handle the diverse and non-standardized layouts of commercial datasheets. As noted in recent studies [18],[19], the structural heterogeneity of datasheets poses a more significant challenge than that of structured academic papers, leaving a robust, unified data extraction framework as an open problem.

Secondly, parameter extraction for the ASM-HEMT model is a significant challenge in itself. The combination of a high-dimensional parameter space and limited data from a single

This work is funded by China Postdoctoral Science Foundation (No. 2022M723913) and Fundamental Research Funds for the Central Universities (No. 3208002309A2). (Corresponding Author: *Hong Cai Chen*)

Yuyang Peng, Jiarui Zhong and Hong Cai Chen are with School of Automation, Southeast University, and also with Ministry of Education Key Laboratory of Measurement and Control of Complex Systems of Engineering, Southeast University, Nanjing 210096, China. (email: pyy990728@sjtu.edu.cn, 220245143@seu.edu.cn, chenhc@seu.edu.cn).

Yang Zhang is with College of Advanced Interdisciplinary Studies, National University of Defense Technology, Changsha, China. (email: 16103271g@connect.polyu.hk).

datasheet creates an ill-posed optimization problem. Various techniques have been developed to address this, evolving from hybrid algorithms [24] to derivative-free optimization (DFO) [25]. More recent machine learning (ML) methods, including neural networks [26] and data augmentation [27], [28], have achieved high accuracy but depend on large training datasets. Consequently, these existing methods are unsuitable for this specific application for two primary reasons. First, conventional optimization algorithms require a well-constrained initial search space that a single datasheet cannot provide. Second, ML models require extensive datasets for training, which contradicts the objective of characterizing a single, specific device. Therefore, a method that can overcome the dual challenges of data sparsity and high-dimensional parameter fitting remains critically needed.

To address these challenges, this paper introduces a novel, end-to-end automated framework for generating ASM-HEMT models directly from PDF datasheets without manual intervention or prior device-specific knowledge. The main contributions of this work are threefold:

**Unified Multi-modal Datasheet Information Extraction Framework:** This framework integrates computer vision and LLMs to robustly parse the complex layouts of datasheets from various manufacturers. It synergistically extracts multi-modal data from text, tables, and graphical curves, ensuring the completeness of the information required for model construction.

**An Iterative-Focusing TPE (IF-TPE) Optimization Algorithm:** This algorithm is designed to implement an efficient parameter extraction under high-dimension and data-insufficient situation. It employs a dual-domain focusing mechanism to adaptively adjust the search space, enabling rapid convergence to high-accuracy solutions even when parameter boundaries are unknown, thereby overcoming the limitations of conventional optimization methods in this scenario.

**The First Complete Automated Workflow from Datasheet to SPICE Model:** By integrating the aforementioned information extraction and parameter optimization modules, we have constructed and implemented the first fully automated, end-to-end workflow from a raw datasheet to a usable SPICE model. We systematically validated this framework on a diverse dataset of 17 commercial HEMT devices from 10 different manufacturers, proving its accuracy, robustness, and practicality, with performance significantly superior to traditional manual methods.

Our proposed method not only liberates engineers from time-consuming and tedious manual operations but also demonstrates exceptional efficiency and accuracy in addressing the challenges of data sparsity. Subsequent chapters will provide a detailed elaboration of the framework's components and present comprehensive experimental results to substantiate its effectiveness.

## II. Fundamental Knowledge

### A. Brief Introduction of ASM-HEMT

The Advanced SPICE Model for High Electron Mobility Transistors (ASM-HEMT) is an industry-standard, physics-based compact model developed specifically for GaN HEMT technologies. Unlike purely empirical models, ASM-HEMT is built upon a surface-potential core, which provides a direct link between the model's parameters and the device's underlying physical processes.[29], [30].

The model's foundation lies in solving fundamental physical equations to accurately determine the charge density of the two-dimensional electron gas (2DEG) at the heterojunction interface. Carrier transport is then described using the drift-diffusion framework. This physics-based approach allows the model to inherently capture the primary device characteristics with high fidelity.

Furthermore, ASM-HEMT is designed to incorporate crucial real-world phenomena that significantly impact performance. These include second-order effects like channel length modulation (CLM), as well as the complex thermal and trapping effects (e.g., current collapse) that are particularly pronounced in GaN devices. By grounding the model in physical principles, ASM-HEMT offers superior scalability and predictive accuracy, making it an essential tool for modern RF and power electronics design.

TABLE I
THE SELECTED SUBSET OF CORE ASM-HEMT PARAMETERS

| Param | description | Param | description |
|---|---|---|---|
| L, W | Channel Length & Width | VSAT | Saturation Velocity |
| LDG, LSG | D/S-Gate Access Length | LAMBDA | Channel Length Modulation |
| tbar | Barrier Thickness | IGSDIO, IGDDIO | G-S/D Diode Sat. Current |
| VOFF | Cut-off voltage | NJGS, NJGD | G-S/D Diode Ideality Factor |
| RHT0 | Thermal Resistance | UTE | Mobility Temp. Dependence |
| NFACTOR | Sub-threshold Slope | DELTA | Exponent for Vdeff |
| U0, UA, UB | Mobility & degradation coeffs | THESAT | Velocity Saturation Parameter |
| EPSILON | AlGaN Permittivity | GDSMIN | Convergence Parameter |
| RSC, RDC | S/D Contact Resistance | IMIN | Minimum Drain Current |
| VSATACCS | Source Access Sat. Velocity | ETA0 | DIBL Parameter |
| VDSCALE | DIBL Scaling VDS | | |
| GAMMA0I, GAMMA1I | Schrodinger-Poisson Params | KTGS, KTGD | G-S/D Current Temp. Coeff. |
| NS0ACCS, NS0ACCD | S/D 2-DEG Density | UTES, UTED | Access Mobility Temp. Dep. |
| U0ACCS, U0ACCD | S/D Access Mobility | MEXPACCS, MEXPACCD | Access Res. Exponent |
| RIGSDIO, RIGDDIO | G-S/D Rev. Current Param | RNJGS, RNJGD | G-S/D Rev. Current Slope |

*1) ASM-HEMT Core*

The complete ASM-HEMT model contains over 200 parameters to comprehensively describe the complex physical behavior of devices. Parameter extraction typically follows a structured, physics-based step-by-step extraction strategy. This strategy relies on a series of specialized electrical characteristic measurements, such as C-V (capacitance-voltage) testing, pulsed I-V (current-voltage) testing, and S-parameter measurements under multiple bias points, to decouple and sequentially determine different parameter subsets within the model [31].

However, this traditional approach faces fundamental challenges in scenarios where only datasheets are available. Commercial datasheets provide only limited, highly integrated performance curves, lacking the detailed underlying measurement data necessary for the aforementioned step-by-step extraction strategy to isolate specific physical effects. This inherent data sparsity renders traditional step-by-step extraction pathways inapplicable, making global optimization the only viable alternative.

Direct global optimization of all 200+ parameters would trigger the curse of dimensionality, leading to unacceptable computational complexity and time costs, while being highly prone to convergence at local optima. Therefore, this study adopts a parameter space reduction strategy. Based on physical significance and correlation with I-V and S-parameters, we carefully selected over 40 core parameters to constitute a dimensionally reduced yet physically complete parameter subset. This approach preserves the model's core predictive capability for key characteristics while dramatically reducing the dimensionality of the optimization problem, establishing a foundation for subsequent efficient and robust parameter extraction using advanced optimization algorithms.

*2) RF Equivalent Circuit for HEMT*

To accurately predict device RF performance in high-frequency regions, extrinsic parasitic effects introduced by pads, leads, and packaging have non-negligible impacts on S-parameters [32]. Therefore, this study integrates a physics-based extrinsic parasitic network [33] on top of the core ASM-HEMT model. The topology of this network represents a classical solution in HEMT small-signal equivalent circuit modeling. As shown in Fig. 1, this network primarily consists of series parasitic resistances ($R_G, R_D, R_S$) and parasitic inductances ($L_G, L_D, L_S$) at the gate, drain, and source, as well as parasitic capacitances ($C_{PG}, C_{PD}$) between gate-ground and drain-ground.

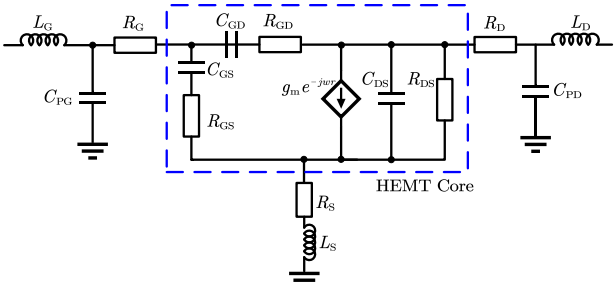

**Fig. 1.** The HEMT small-signal equivalent circuit, including the intrinsic core model and the extrinsic parasitic network.

The subsequent parameter extraction objective of this work is to accurately determine the values of these extrinsic parasitic parameters and core intrinsic parameters, thereby ensuring that the complete hybrid model can precisely reproduce S-parameter characteristics across a wide frequency range.

*B. Typical Datasheets Format*

The semiconductor device datasheet is a foundational technical document issued by manufacturers. It provides a comprehensive specification of the device's electrical performance, operational parameters, and physical attributes, serving as the authoritative reference for circuit design and device modeling.

However, datasheets pose significant challenges for automated information extraction due to their multi-modal, unstructured nature. These documents typically combine textual data, graphical elements, and complex layouts within PDF formats (Fig. 2), creating barriers for systematic data parsing. Critical information is distributed across multiple heterogeneous formats, including unstructured text, parameter tables with varying formats, circuit schematics, and characteristic curve plots without provided raw data points. Particularly, these curves encode the fundamental behavioral characteristics of devices in graphical form. Their accurate digitization constitutes both the most crucial and technically challenging step in subsequent device modeling workflows.

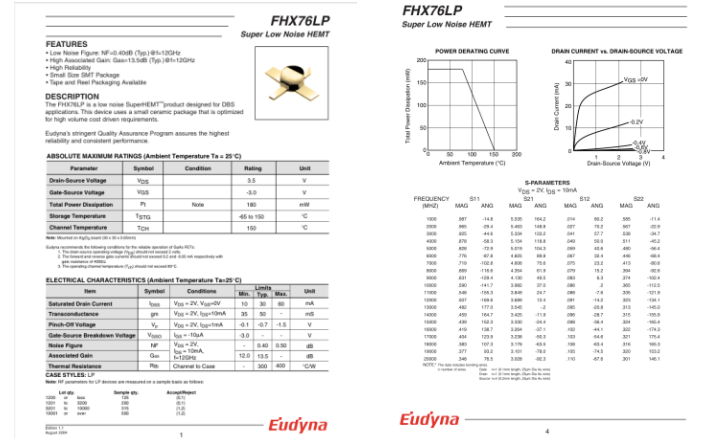

**Fig. 2.** typical datasheet for a commercial HEMT, demonstrating the multi-modal layout that mixes texts, parameter tables, and characteristic plots.

When conducting parameter extraction for complex compact models like ASM-HEMT, key information in datasheets can be categorized into two types:

**Indicative Parameters**: This category primarily includes typical values or maximum/minimum values in tables, such as threshold voltage $V_{GS(th)}$, saturation drain current $I_{DSS}$, on-resistance $R_{DS(on)}$, etc. Under the "no prior knowledge" framework proposed in this study, these discrete numerical points are not used to guide or constrain the initial parameter search range. Instead, our method relies solely on the officially defined parameter list of ASM-HEMT to establish a unified, extremely broad initial search space. Therefore, the core value of these indicative parameters lies in serving as critical, independent validation benchmarks after completion of the parameter extraction process, to verify whether the final generated model is consistent with the device's nominal behavior in key performance metrics.

**Extraction-Critical Data**: This constitutes the core data driving model parameter optimization. For ASM-HEMT models, it primarily originates from the following two types of performance curve plots:

- **I-V Characteristic Curves**: Output characteristic curves ($I_D$ vs. $V_{DS}$) and transfer characteristic curves ($I_D$ vs. $V_{GS}$) contain DC and quasi-static behavioral information of devices under different bias conditions, serving as the foundation for extracting core parameters related to carrier transport, channel charge control, and other phenomena in the model.
- **S-Parameters**: S-parameters presented in tabular or Smith chart formats serve as the sole basis for extracting parameters related to high-frequency dynamics and extrinsic parasitic effects (such as parasitic capacitances and inductances) in the model.

While, extracting accurate data from characteristic curves and S-parameters is non-trivial, requiring specialized methodologies and model-specific processing to ensure reliable parameter extraction.

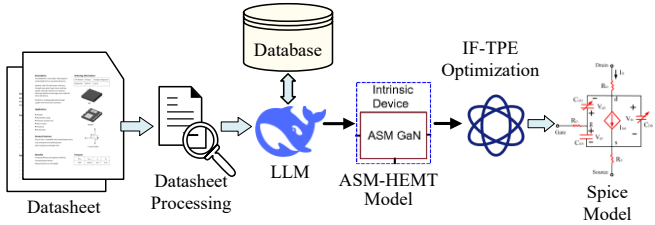

**Fig. 3.** The proposed end-to-end framework for automatically generating SPICE models from datasheets using LLMs and the IF-TPE optimization algorithm.

## III. THE PROPOSED FRAMEWORK

This paper proposes a fully automated, end-to-end HEMT parameter extraction framework that requires no human intervention, with its overall architecture shown in Fig. 3. This workflow aims to efficiently transform raw, unstructured datasheet PDF documents into accurate, usable SPICE models. The entire framework integrates advanced artificial intelligence technologies to decompose the complex modeling task into two core stages: Data Processing and Extraction, and Parameter Optimization.

The workflow begins with the **Datasheet Parsing** stage. In this stage, a dedicated LLM first performs intelligent parsing of the input PDF document, automatically identifying and separating multimodal information including text, tables, and graphical curves. Subsequently, an integrated computer vision module performs high-precision digitization of graphical content, converting visualized characteristic curves into numerical datasets for subsequent use. All extracted and structured multimodal data collectively constitute a comprehensive device characteristic database.

In the **Parameter Optimization** stage, the extracted data (primarily I-V and S-parameter curves) serve as optimization targets. Our proposed IF-TPE algorithm performs global optimization directly within a predefined, physics-constrained high-dimensional parameter space. Through iterative search, this algorithm seeks the optimal parameter set that minimizes the error between model simulation results and datasheet-extracted data. Ultimately, this framework can automatically generate a high-precision ASM-HEMT model that accurately characterizes the device behavior described in the original datasheet.

*A. Datasheet Processing*

To automatically extract the required I-V characteristic curves and S-parameter tables from target documents, this framework employs a three-stage automated data extraction process, as shown in Fig. 4.

Firstly, the datasheet PDF document is first parsed using a layout analysis model EDocNet [23]. This model identifies bounding boxes and categories of various elements on each page to understand the document's detailed layout. The datasheet is decomposed into fundamental layout elements such as text blocks, headings, tables, and images. Secondly, a LLM model, Deepseek [34], is applied to semantically classify all parsed image and table elements. It identifies target I-V characteristic curves and S-parameter tables among these elements. Finally, the identified tables and images of curves are further processed using specialized models to obtain numerical data values, enabling subsequent analysis.

Deepseek lacks native PDF preprocessing tools, relying instead on tools like pdfplumber [35]. Since these tools are not designed for datasheets, they handle datasheet layouts poorly, often losing information when used directly. Integrating our in-house EDocNet for preprocessing significantly boosts parsing capability.

*1) Automated S-Parameter Extraction*

To automate the extraction of S-parameter data required for RF circuit construction, we've developed a workflow that processes identified tables using EDocNet and Deepseek. First, EDocNet parses the internal text of detected tables, structuring them into machine-readable formats. Next, Deepseek extracts specific parameters from this structured text, including bias conditions ($V_{DS}$, $I_D$) and complete frequency-dependent S-parameter data points. This unified approach ensures efficient retrieval of all necessary data for RF model calibration and validation.

*2) Automated Curve Extraction Model EDocCurve*

For the recognized IV curve charts, EDocCurve is developed to convert them into numerical data points to enable subsequent analysis.

First, the framework uses the YOLO object detection model to locate bounding boxes for all axis scale labels and parameter labels in the image. Optical Character Recognition (OCR) is applied to image slices within these bounding boxes to extract corresponding numerical values or parameter texts. Each successfully identified scale value, paired with its pixel coordinates in the image, forms a calibration point. Meanwhile, Hough Transform precisely detects orthogonal coordinate axes, whose intersection serves as the origin of the pixel coordinate system. The output of this step is a set of calibration points for each axis, containing both physical coordinates ($V_{\text{real}}$) and pixel coordinates ($p_{\text{pixel}}$).

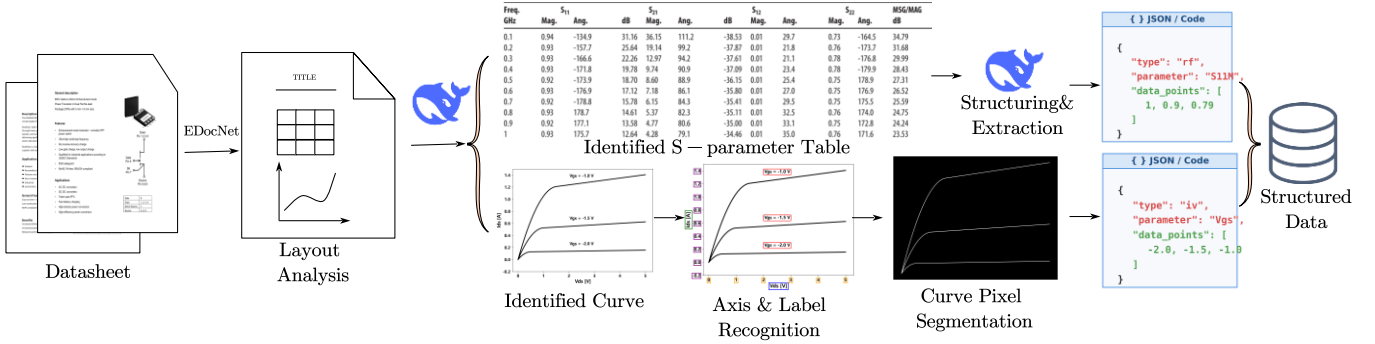

**Fig. 4.** The data extraction and structuring pipeline. An LLM classifies layout elements parsed by EDocNet, branching the process to handle tables (using Deepseek) and curves (using EDocCurve) separately to generate structured data.

Secondly, the framework uses an image segmentation method to directly identify and output pixel sequences that form data curves. Given a chart image, the model effectively handles complex scenarios such as curve intersections and noise, directly outputting one or more high-precision pixel coordinate sets $\{S_k\}$, where each set represents an independent curve trajectory, laying the groundwork for subsequent data conversion.

Thirdly, the mapping of pixel coordinates ($p_{\text{pixel}}$) to physical coordinates ($V_{\text{real}}$) is achieved by establishing and solving an affine transformation model. For linear axes, the relationship is $V_{\text{real}} = \alpha \cdot p_{\text{pixel}} + \beta$; for logarithmic axes, it is $V_{\text{real}} = 10^{\alpha \cdot p_{\text{pixel}} + \beta}$, where $\alpha$ (scaling factor) and $\beta$ (offset) are parameters to be determined. Using the axis calibration points obtained in the first step, a system of linear equations is constructed and solved to precisely determine the transformation parameters for each axis independently. This allows any pixel coordinate to be mapped to its corresponding physically meaningful real coordinate via the transformation model.

Finally, after determine the mapping between axis and pixels, the values of curves can be extracted. Since each characteristic curve chart contains multiple curves, the framework matches each extracted I-V curve to its corresponding gate voltage ($V_{\text{gs}}$) parameter. It uses YOLO and OCR to identify curve parameter labels (e.g., "$V_{\text{gs}}$=1.0V") in the image, then associates each label with the nearest curve pixel set in spatial position. After matching, all pixel points on each curve are converted to real ($V_{\text{ds}}, I_{\text{ds}}$) coordinate pairs via the established affine transformation model. These pairs are combined with the associated $V_{\text{gs}}$ parameter to generate complete 3D data points ($V_{\text{ds}}, I_{\text{ds}}, V_{\text{gs}}$).

*B. Iterative Focusing TPE for Parameter Extraction*

*1) Optimization Problem Definition:*

ASM-HEMT parameter extraction fundamentally constitutes a high-dimensional nonlinear optimization problem that seeks optimal parameter combinations to achieve the best match between simulation results and experimental data. The optimization problem can be formally defined as follows:

Decision Variables $\theta$: The ASM-HEMT parameter vector to be extracted is defined as $\theta = [\text{param}_1, \text{param}_2, ..., \text{param}_N]$, where N represents the total number of model parameters.

Search Space $\Omega$: Hard constraint boundaries are defined based on the physical significance of parameters, embodying the key characteristic of "prior-knowledge-free" approach—leveraging universal semiconductor physics knowledge rather than device-specific prior information.

Objective Function $L(\theta)$: Following the standard root-mean-square error calculation methodology, the objective function is defined as:

$$L(\theta) = \text{RMSE}(\text{Sim}(\theta), \text{Data}) \qquad (1)$$

In practical implementation, I-V characteristics and S-parameters are optimized separately to ensure optimal fitting accuracy for each domain.

For I-V characteristic optimization:

$$\text{RMSE}_{\text{iv}} = \sqrt{\frac{1}{M}\sum_{i=1}^{M}[I_{\text{sim}}(V_{\text{gs},i}, V_{\text{ds},i}; \theta) - I_{\text{exp}}(V_{\text{gs},i}, V_{\text{ds},i})]} \qquad (2)$$

For S-parameter optimization:

$$\text{RMSE}_{\text{s}} = \sqrt{\frac{1}{K}\sum_{i=1}^{K}\|S_{\text{sim}}(f_k; \theta) - S_{\text{exp}}(f_k)\|} \qquad (3)$$

where $M$ and $K$ represent the number of I-V test points and frequency points, respectively, and $\|.\|$ denotes the complex modulus.

This optimization problem definition provides a clear mathematical framework for the subsequent TPE Bayesian optimization algorithm, enabling the automated parameter extraction process to efficiently converge to the global optimum under reasonable physical constraints.

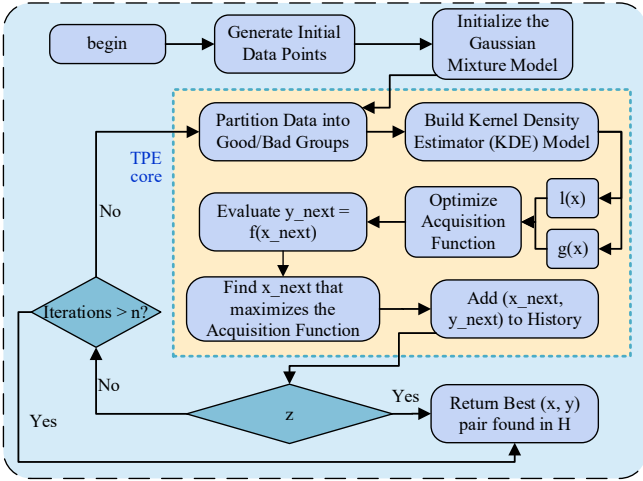

**Fig. 5.** Flowchart of the Tree-structured Parzen Estimator (TPE) algorithm.

*2) Tree-structured Parzen Estimator (TPE)*

TPE is a tree-structured Bayesian optimization method designed to solve global optimization problems for black-box functions (Fig. 5). In each trial, for each parameter, TPE maintains a Gaussian mixture model $l(x)$ for parameters associated with the best objective values according to a predefined proportion, and maintains another Gaussian mixture model $g(x)$ for the remaining parameters. The hyperparameters corresponding to the maximization of $l(x)/g(x)$ are selected as the next set of search values. Through this approach, the TPE algorithm can adaptively adjust the size of the parameter search space and identify the global optimum within the fewest possible iterations.

$$p(x|y) = \begin{cases} l(x) & \text{if } y < y^* \\ g(x) & \text{if } y > y^* \end{cases} \quad (4)$$

where $l(x)$ is constructed using the observation space $\{x_k\}$ corresponding to $f(x_k) < y^*$, while the remaining observation space with $f(x_k) > y^*$ is used to construct $g(x)$.

TPE selects expected improvement as the acquisition function. Since $p(y|x)$ cannot be directly obtained through observation, Bayesian formula is employed for transformation, yielding:

$$\begin{aligned} \mathrm{EI}_{y^*}(x) &= \int_{-\infty}^{y^*} (y^* - y)\, p(y|x)\, dy \\ &= \int_{-\infty}^{y^*} (y^* - y)\, \frac{p(x|y)\, p(y)}{p(x)}\, dy \end{aligned} \quad (5)$$

where $y^*$ represents the threshold.

Let $\gamma = p(y < y^*)$ then according to the Law of Total Probability and marginalization, we obtain:

$$p(x) = \int p(x|y)\, p(y)\, dy = \gamma l(x) + (1-\gamma) g(x) \quad (6)$$

Finally, EI can be simplified to:

$$\mathrm{EI}_{y^*}(x) = \frac{\gamma y^* l(x) - l(x)\int_{-\infty}^{y^*} p(y)\, dy}{\gamma l(x) + (1-\gamma) g(x)} \propto 1 \Big/ \Big[(1-\gamma)\frac{g(x)}{l(x)} + \gamma\Big] \quad (7)$$

The above equation indicates that to maximize EI, our optimization objective is to achieve high probability under $l(x)$ while maintaining low probability under $g(x)$ at point x. In each iteration, the algorithm returns the candidate $x^*$ with the maximum EI.

*3) The Iterative Focusing TPE Algorithm (IF-TPE)*

Although the standard TPE algorithm provides an efficient framework for black-box optimization, it still faces challenges when directly applied to the specific problem of extracting ASM-HEMT parameters. The primary reason is that under the "no prior knowledge" setting, the initial search space Ω is extremely broad, and the effective data points provided by datasheets are relatively sparse, making it difficult for the standard TPE's surrogate model to quickly and accurately capture the distribution patterns of optimal parameters in the early stages of optimization.

To address this issue, this paper proposes an enhanced Bayesian optimization algorithm called Iterative Focusing TPE (IF-TPE). The core idea of this algorithm is to introduce a batch-wise, dual-domain focusing adaptive strategy that intelligently guides TPE's search process in both the Spatial Domain and Resolution Domain.

The complete workflow of IF-TPE is shown in Fig. 6(a), which decouples the optimization process into a nested dual-loop structure: the outer loop is responsible for strategically updating the search space, while the inner loop performs efficient parameter exploration within the currently focused window.

**(1) Spatial Focusing: Adaptive Update of the Search Window**

The spatial focusing mechanism aims to dynamically relocate and contract the search window to new high-potential regions based on existing optimal solutions. In the $k$-th batch, the search space is defined as a dynamic hyperrectangular window $\Omega_k$, which is determined by its center vector $c_k$ and range vector $r_k$.

$$\Omega_k = \{\theta \mid c_k - r_k \leqslant \theta \leqslant c_k + r_k\} \quad (8)$$

Upon the completion of a batch of TPE iterations, the algorithm updates the search space for the subsequent batch, $\Omega_{k+1}$, based on the optimal parameter solution, $\theta_k^*$, identified within that batch. This process corresponds to the operation performed by the "Spatial & Precision Focusing" module depicted in Fig. 6(a).

**Center Vector Update**: To concentrate subsequent searches around the currently known optimal solution, the new search center $c_{k+1}$ is directly set to the optimal parameter $\theta_k^*$ from the previous batch. This strategy ensures that the optimization process can continuously explore in the direction of the most significant performance improvement:

$$c_{k+1} = \theta_k^* \quad (9)$$

**Range Vector Update**:

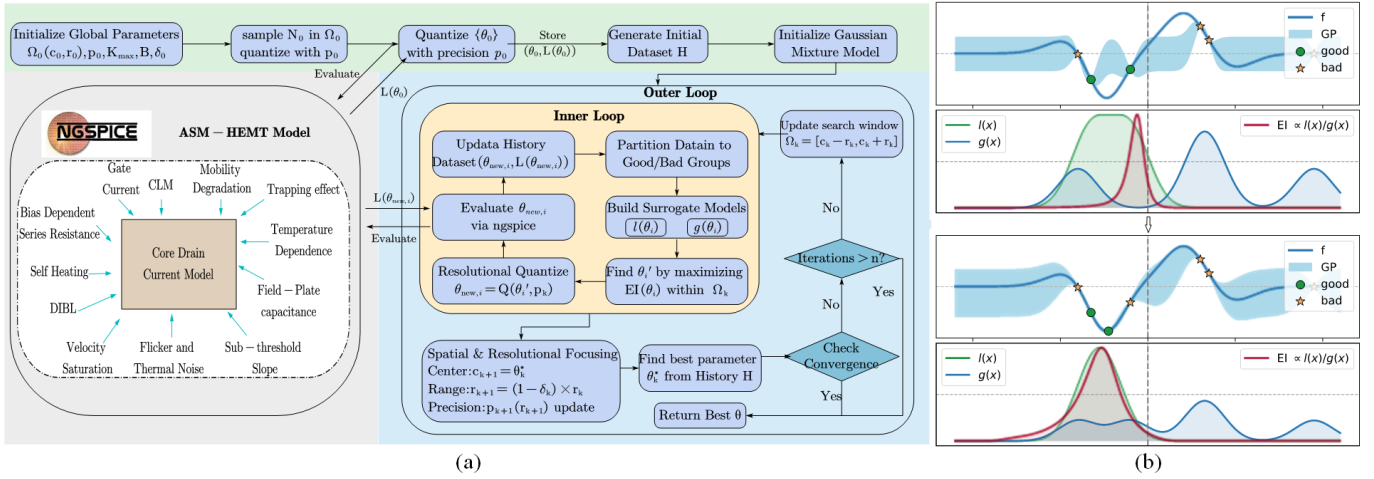

**Fig. 6.** The flowchart of the proposed IF-TPE process for parameter extraction of the ASM-HEMT model. (a) The dual-loop flowchart for adaptive parameter extraction. (b) Illustration of the TPE process, where the Expected Improvement (EI) acquisition function, derived from models of good ($l(\theta)$) and bad ($g(\theta)$) points, guides the optimization.

As optimization progresses, the search range needs to gradually contract to achieve fine-tuning. We introduce a contraction rate $\delta_k$ that decays with batch k to realize a smooth transition from global exploration to local optimization.

$$r_{k+1} = r_k \cdot (1 - \delta_k) \quad (10)$$

where the decay contraction rate $\delta_k$ can be modeled as an exponential decay function, ensuring that the algorithm contracts rapidly in the early stages while being more robust in later stages to prevent premature convergence:

$$\delta_k = \delta_{init} \cdot e^{-\alpha \cdot k} \quad (11)$$

Here, $\delta_{init}$ is the initial contraction rate, and $\alpha$ is the decay constant.

**(2) Resolutional Focusing: Adaptive Precision Quantization**

To prevent floating-point computations within the vast parameter space at a precision below the effective physical limit of the device, IF-TPE incorporates an adaptive precision quantization mechanism. This mechanism dynamically adjusts the discretization step size explored by the optimizer in each parameter dimension according to the scale of the current search window. This procedure corresponds to the "Resolutional Quantize" module in the workflow depicted in Fig. 6(a).

For the j-th dimensional parameter, the search range during the k-th batch is given by $[\alpha_k^{(j)}, \beta_k^{(j)}] = [c_k^{(j)} - r_k^{(j)}, c_k^{(j)} + r_k^{(j)}]$. Its discretization step size $s_k^{(j)}$ is defined as follows:

$$s_k^{(j)} = \frac{\beta_k^{(j)} - \alpha_k^{(j)}}{N_{grid}} \quad (12)$$

where $N_{grid}$ is a predefined hyperparameter representing the number of grid points divided within the current one-dimensional search interval.

The novelty of this design lies in the fact that while the number of grid points, $N_{grid}$, is fixed, the absolute step size, $s_k^{(j)}$, is adaptive. In accordance with the aforementioned spatial focusing mechanism, the search range $\beta_k^{(j)} - \alpha_k^{(j)}$ progressively contracts as the batch number, k, increases; consequently, the step size $s_k^{(j)}$) decreases automatically. This mechanism ensures that the search resolution is automatically refined as the spatial window shrinks, enabling a seamless transition from coarse-grained exploration to fine-grained optimization. This, in turn, facilitates the rapid and accurate extraction of parameters within a limited simulation budget. The complete process of IF-TPE can be formally summarized through Algorithm 1.

---

**Algorithm 1:** The Iterative Focusing TPE Algorithm

// 1. Initialization
Initialize $c_0$, $r_0$, $p_0$
Sample $N_0$ points $\{\theta_0\}$ in $\Omega_0 = [c_0 - r_0, c_0 + r_0]$
Quantize $\{\theta_0\}$ with precision $p_0$
Evaluate all $L\{\theta_0\}$ via circuit simulation.
Initialize history dataset $H \leftarrow \{(\theta_0, L(\theta_0))\}$
// 2. Iterative Optimization
for $k = 0$ to $K_{max} - 1$ do
    $\Omega_k = [c_k - r_k, c_k + r_k]$
    for $i = 1$ to $B$ do
        Partition H into "good" and "bad" sets.
        Build KDE models $l(\theta_i)$ and $g(\theta_i)$.
        $\theta_i' \leftarrow \arg\max_{\theta_i \in \Omega_i} EI(\theta_i)$
        $\theta_{new,i} \leftarrow Q(\theta_i', p_k)$
        $L_{new,i} \leftarrow L(\theta_{new,i})$ via circuit simulation.
        $H \leftarrow H \cup \{(\theta_{new,i}, L_{new,i})\}$
        Build KDE models $l(\theta_i)$ and $g(\theta_i)$
    end for
    $\theta_k^* \leftarrow \arg\min_{\theta \in H} L(\theta)$
    Update center: $c_{k+1} \leftarrow \theta_k^*$
    Update range: $r_{k+1} \leftarrow (1 - \delta_k) \cdot r_k$.
    Update precision $p_{k+1}$ based on $r_{k+1}$.
end for
$\theta_{best} \leftarrow \arg\min_{\theta \in H} L(\theta)$
return $\theta_{best}$

## IV. Experimental Setup and Results

In the validation, all the simulations were performed on a computer equipped with an Intel Core i9-13900K CPU and 64GB DDR5 RAM, running on Windows 11 pro. The ngspice simulator (version 42) was used with the ASM-HEMT v101.4.0 device model.

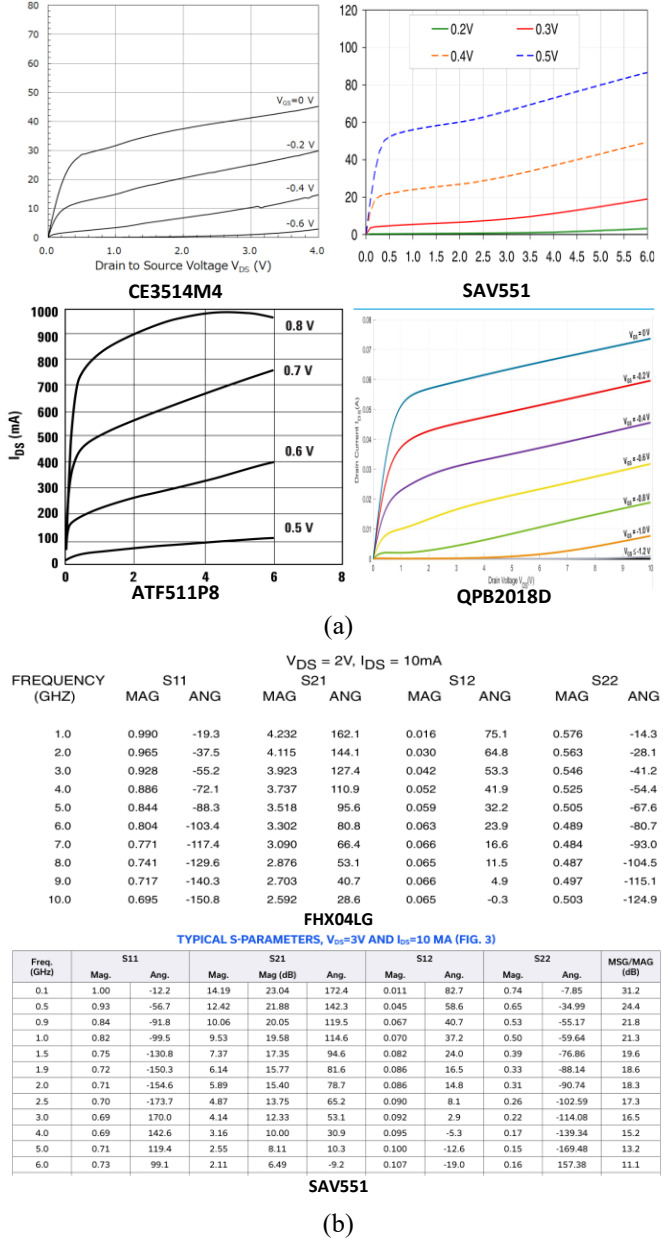

**Fig. 7.** A collage of screenshots from the datasheets of selected devices in the dataset, highlighting the visual and structural diversity of their characteristic (I-V) curves and S-parameter tables.

### A. Dataset Description

To comprehensively validate the effectiveness and robustness of the proposed framework, this study established a comprehensive dataset of commercial HEMT datasheets. This dataset comprises 17 devices from 10 leading semiconductor manufacturers, covering a wide spectrum of types from GaN power HEMTs (e.g., Innoscience INN650DA04, RealChip RC65E300Y) to RF/microwave pHEMTs (e.g., Qorvo QPD2018D, Mini-Circuits SAV-551+).

The core feature of this dataset is its high degree of heterogeneity, which provides a robust basis for evaluating the framework's performance under realistic and complex scenarios. This heterogeneity is primarily manifested in the following aspects:

**Diversity of Device Types**: The dataset covers devices ranging from 650V GaN HEMTs for high-power electronic applications (total power dissipation from 38W to 55.5W) to low-noise pHEMTs for RF/microwave communication (output power at 1-dB compression point from 12.0 dBm to 31 dBm).

**Diversity of Package Formats**: The collection includes both surface-mount packages (e.g., DFN, SOT-343, LPCC) suitable for automated manufacturing and bare die formats intended for hybrid integrated circuits.

**Diversity of Datasheet Layouts**: A core challenge for the framework is the profound diversity in datasheet layouts across different manufacturers, as illustrated in Fig. 7. These inconsistencies manifest in both graphical and tabular data representations. For graphical elements like characteristic curves, variations include fundamental presentation styles (e.g., color vs. monochrome), line types (e.g., solid vs. dashed), and disparate axis scales and ranges (Fig. 7 (a)). For tabular data, structures for presenting parameters, such as S-parameters, are similarly inconsistent (Fig. 7 (b)). This heterogeneity necessitates a highly robust and adaptable data extraction methodology.

This diversity in formatting mandates that the automated extraction framework possess exceptional robustness and adaptability, precluding any reliance on hard-coded rules tailored to specific templates. A primary objective of this framework is, therefore, to automatically and accurately extract numerical data suitable for model parameter optimization from these images and tables of disparate formats.

### B. Datasheet Information Extraction Results

To validate the effectiveness of the proposed framework, we firstly performed an end-to-end parameter extraction procedure on the dataset. Fig. 8 illustrates a typical example of this process, where the system is processing a page from a device datasheet. As shown, our layout analysis module successfully performs semantic segmentation on the page, accurately identifying and localizing key information regions of different modalities. This precise content localization is a critical prerequisite for achieving subsequent automated extraction. It enables the framework to direct the identified specific regions (such as figures or tables) to the corresponding specialized extraction modules for in-depth processing.

To present the effectiveness of the EDocCurve framework, we use the commercial pHEMT device ATF38143 as a case study and its complete curve extraction workflow is illustrated in Fig. 9. The workflow commences with the original datasheet image shown in Fig. 9(a). First, a YOLO-based key element detection module (Fig. 9(b)) localizes and identifies all axes, ticks, and parameter labels, and utilizes OCR to extract their textual values for subsequent coordinate calibration. Subsequently, an image segmentation model separates the curve trajectories from the background, generating a high-precision binary mask as depicted in Fig. 9(c). To further refine and vectorize the mask pixels, the framework integrates a line-smooth model to handle potential discontinuities. Finally, through an affine transformation model, we map the segmented curve pixels to the physical coordinate system calibrated by the OCR results and match the parameter labels to the nearest curves. The final reconstructed result, shown in Fig. 9(d), exhibits high fidelity with the original image (Fig. 9 (a)), thereby validating the end-to-end accuracy and robustness of the EDocCurve framework.

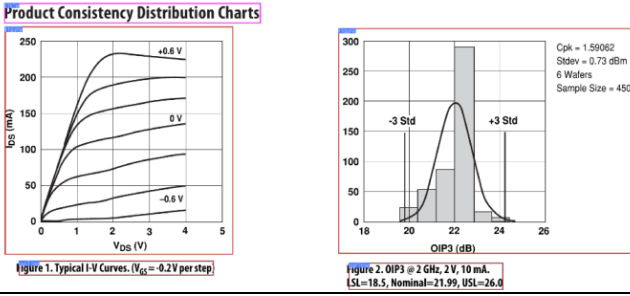

**Fig.10.** Automated S-parameter table extraction, on MGFC4419G. (a) The original table from the datasheet. (b) The extracted data structured into a machine-readable format.

Fig.10 illustrates our extraction result of tabular S-parameter data using Deepseek. Our system accurately identifies and parses the S-parameter table within the datasheet (Fig.10 (a)). This process not only extracts the frequency-dependent S-parameter values (such as the magnitude and phase of S11) but also simultaneously identifies the key bias conditions defined in the table's title or notes. As shown in Fig.10(b), all extracted information is automatically converted into a unified, machine-readable, structured format. This capability ensures the integrity and accuracy of the data required for subsequent small-signal model parameter fitting.

In summary, through the automated workflow proposed by this framework, the I-V curves and S-parameter data for all devices in our test dataset were successfully extracted. These high-fidelity numerical results are uniformly stored in a structured database, laying a solid data foundation for subsequent batch-processed model parameter extraction and validation.

*C. ASM-HEMT Parameter Extraction Results*

To comprehensively validate the performance of the proposed automated framework in terms of accuracy, efficiency, and versatility, we executed an end-to-end parameter extraction procedure on the entire dataset covering 17 devices. Notably, only four of them provide S-parameters.

Table II provides a quantitative summary of the automated extraction results for all devices. As indicated in the "Doc Suc" (Document Processing Success Rate) column, the framework successfully parsed all 17 datasheets from different manufacturers with various layouts. The I-V RMSE and S-RMSE columns present the final normalized root-mean-square errors for the DC and RF parameter optimization, respectively, demonstrating the high accuracy of the model. Notably, driven by the proposed IF-TPE algorithm, the parameter optimization tasks for most devices converged in fewer than 500 iterations, which fully demonstrates the excellent optimization efficiency of our framework.

**Fig. 8.** Results of the layout analysis for a datasheet page using EDocNet.

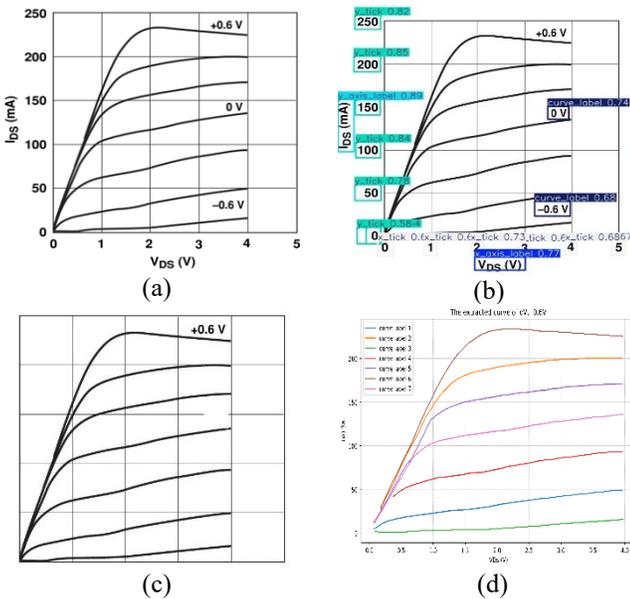

**Fig. 9.** The multi-stage pipeline of the EDocCurve model, demonstrated on the ATF38143 pHEMT I-V characteristics.

TABEL II
PARAMETER EXTRACTION ON A DIVERSE HEMT DATASET

| Device Model | Doc Suc. | IV-RMSE(%) | S-RMSE(%) | Iteration |
|---|---|---|---|---|
| ATF511P8 | √ | 4.85 | \ | 440 |
| ATF38143 | √ | 2.23 | \ | 280 |
| CE3514M4 | √ | 2.24 | \ | 400 |
| CE3521M4 | √ | 1.56 | \ | 240 |
| EPC2037 | √ | 2.37 | \ | 900 |
| FHX04LG | √ | 1.77 | 18.7 | 540 |
| FHX04X | √ | 1.95 | 9.09 | 500 |
| FHX13LG | √ | 2.31 | 14.5 | 500 |
| FHX13X | √ | 2.08 | 9.06 | 740 |
| INN650DA04 | √ | 2.51 | \ | 280 |
| MGFC4419G | √ | 1.17 | \ | 280 |
| QPD2018D | √ | 2.85 | \ | 320 |
| QPD2025D | √ | 2.27 | \ | 500 |
| QPD2120D | √ | 3.11 | \ | 400 |
| RC65E300Y | √ | 1.43 | \ | 320 |
| SAV551 | √ | 3.18 | \ | 700 |
| TAV581 | √ | 2.57 | \ | 500 |

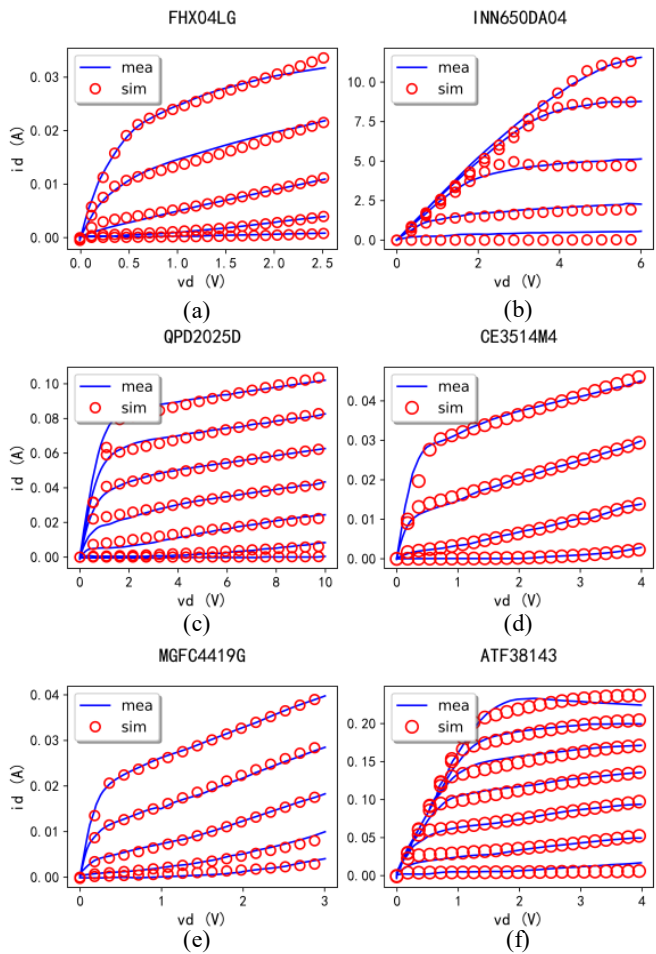

**Fig. 11.** Fitting results of the DC output characteristics (ID−VDS curves) for six representative HEMT devices.

Fig. 11 provides a visual validation of the fitting results for the DC output characteristics (ID−VDS curves) of 6 representative devices. In the plots, the red circles (mea) represent the measured data automatically digitized from the datasheets, while the solid blue lines (sim) denote the simulated model curves obtained using the parameters extracted by our framework.

As shown in the figure, the fitting results exhibit a high degree of consistency across the entire dataset. The model accurately reproduces the DC behavior under various gate voltages, both for the high-current GaN power HEMT in (b) (INN650DA04) and for the low-power RF pHEMTs from different process technologies in (a) and (c)-(f). The simulation curves accurately capture the characteristics of the device's linear region and saturation region, demonstrating the framework's strong adaptability and high fidelity across different device types and operating ranges.

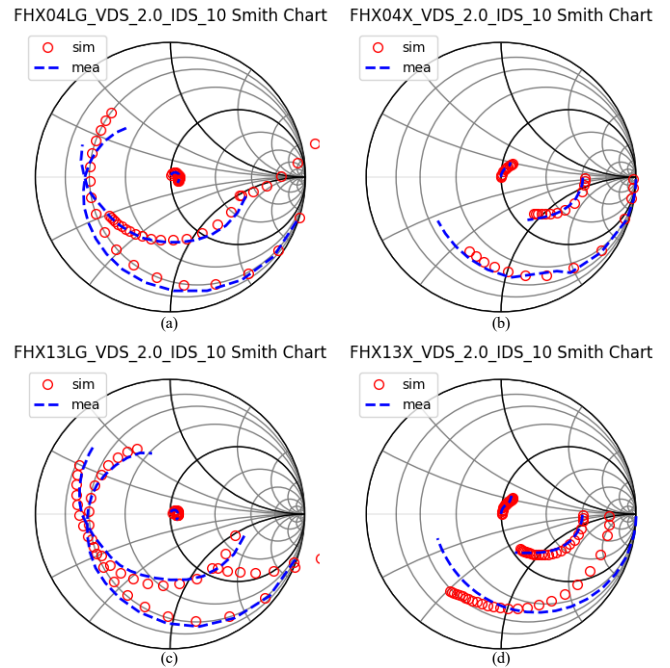

**Fig. 12.** Comparison of simulated (sim) vs. measured (mea) S-parameter results on the Smith chart for four FHX-series RF HEMTs.

To validate the model's accuracy in the high-frequency regime, we also modeled and compared the S-parameters of the RF devices. Fig. 12 displays a comparison of the simulated and measured S-parameter results for four FHX-series devices on a Smith Chart.

As illustrated in the figure, the model simulation results (sim, red circles) exhibit excellent consistency with the measured data extracted from the datasheets (mea, blue dashed lines) across the entire frequency sweep range from low to high frequencies. This high degree of visual agreement is supported by the very low S-RMSE quantitative metrics presented in Table II. The precise fitting of the S-parameters demonstrates that our framework, by optimizing the intrinsic ASM-HEMT model and the extrinsic parasitic network, can effectively capture the complex high-frequency dynamic characteristics of the device, which is crucial for the design of RF amplifiers and communication systems.

In summary, the dual validation results of DC and RF characteristics strongly demonstrate that the proposed automated parameter extraction framework possesses high accuracy and reliability, successfully transforming unstructured datasheet information into precise, usable SPICE models.

## V. CONCLUSION

In this work, we propose and validate a novel, end-to-end automated framework capable of generating high-fidelity ASM-HEMT SPICE models directly from PDF datasheets. The framework's success stems from two core innovations: a unified multi-modal information extraction module that robustly parses heterogeneous datasheet layouts by combining computer vision and Large Language Models (LLMs), and a novel Iterative-Focusing TPE (IF-TPE) optimization algorithm that effectively overcomes the challenge of sparse-data-driven parameter extraction. By integrating these components, we established the first fully automated workflow from datasheet to SPICE model. Systematic validation on a comprehensive dataset of 17 commercial HEMT devices from 10 different manufacturers demonstrates that the generated models achieve excellent agreement with the original datasheet characteristics, thereby confirming the framework's accuracy, robustness, and practical utility.